\documentclass[3p,times,twocolumn]{article}
\usepackage{braket}
\usepackage{graphicx}
\usepackage{amsmath}
\usepackage{here}
\usepackage[hidelinks]{hyperref}
\usepackage{authblk}

\usepackage{xcolor}

\newcommand{\ketbrac}[2] { {\text{$\mathfrak{C}_{{#1}{#2}}$}} }
\newcommand{\projector}[2]{{\text{$\mathbb{P}^{({#1})}_{{#2}}$}}}
\newcommand{\sets}[1]{{\text{$\mathfrak{s}^{\dagger}_{{#1}}$}}}
\newcommand{\scraps}[1]{{\text{$\mathfrak{s}_{{#1}}$}}}

\newcommand{\innerid} {{\text{$\mathfrak{i}$}}}

\newcommand{\bmat}{\left[\begin{array}}
\newcommand{\emat}{\end{array}\right]}

\usepackage{amssymb}

\usepackage[figuresright]{rotating}

\title{
Towards few-body QCD on a quantum computer \\
{\small Proceedings of the 27th High-Energy Physics International Conference in Quantum Chromodynamis (QCD24)}} 

\author{J.J. Gálvez-Viruet}

   \affil{Theoretical Physics Department \& IPARCOS, Universidad Complutense de Madrid, Pl. de las Ciencias 1 - 28040 Madrid
}

\begin{document}
\maketitle
\begin{abstract}
\noindent
Quantum computers are promising tools for the simulation of many-body systems, and among those, QCD stands out by its rich phenomenology. Every simulation starts with a codification, and here we succently review a newly developed compact encoding based on the identification between registers and particles; the quantum memory is divided into registers, and to each we associate a Hilbert space of dimension the number of degrees of freedom of the codified particles. In this way we gain an exponential compression over direct encodings for a low number of particles with many degrees of freedom. As an example we apply this encoding on a two-register memory and implement antisymmetrization and exponentiation algorithms.

\end{abstract}

\section{Introduction}
Numerical simulations of QCD for the description of its phase space \cite{de_forcrand_1006_2010} or the evaluation of real-time observables are thwarted by the so-called sign problem of Monte Carlo sampling \cite{de_forcrand_1006_2010,ortiz_0012334_2001}, which makes it challenging, for example, to derive an EoS for QCD matter in neutron stars' cores; where only loose bounds by pQCD and chiral EFT \cite{hammer_2006_2020, llanes-estrada_1911_2019,lope-oter_2409_2022} can be derived. 

Quantum computers (QC), by their very definition, can realize Hamiltonian evolution on quantum memories, usually based on two level systems called qubits \cite{lloyd_009608_1996,nielsen_quantum_2012}. They are thus envisioned as powerful tools to reach there where classical computers cannot \cite{feynman_simulating_1982}. Their first applications as simulators were centered on fermion systems \cite{ortiz_0012334_2001,abrams_9711_1997} and soon reached quantum chemistry \cite{whitfield_1103_2011,mcardle_2003_2020} and HEP \cite{bauer_2204_2022}, with implementations based on lattice formulations \cite{jordan_1206_2012,banuls_1911_2020,farrell_preparations_2023} and works that explore other implementations \cite{barata_201200020_2021,kirby_210510941_2021,qian_2212_2022,kreshchuk_2203_2022,chawdhry_2311_2023,galvez-viruet_2409_2024}. Among their most recent applications we find studies for the simulation of nuclear systems \cite{perez-obiol_2307_2023,wang_2406_2024}, heavy quarks \cite{gallimore_2304_2023,de_arenaza_2409_2024}, light mesons \cite{qian_2212_2022}, in medium jets \cite{barata_220806750v1_2022,barata_230701792v1_2023} and thermalization and phase diagrams \cite{motta_2002_2020,zhou_2207_2022,davoudi_2308_2023}, among others.

Simulating a system requires encoding its degrees of freedom (quantum numbers). The ``direct'' encodings, based on the identification of the occupation number basis and the computational basis are the most common, and they generally need a qubit for each \textit{fermionic mode} to simulate; the Jordan-Wigner \cite{jordan_uber_1928} and the Bravyi-Kitaev \cite{bravyi_fermionic_2002} encodings are two prominent examples. Here we present an implementation based on a ``compact encoding'' scheme \cite{galvez-viruet_2409_2024} in which the memory is divided into registers with enough size to accommodate the basis states of each particle. This codification is then used for the implementation of time evolution under a Hamiltonian with kinetic and exchange terms. This encoding is expected to be efficient for studying few-body QCD problems, given the large number of internal degrees of freedom associated with quarks and gluons.
\section{Codification}
We now review the particle-register codification discussed in \cite{galvez-viruet_2409_2024}. Consider a quantum memory of qubits divided into registers, each representing a particle with spin and momentum,
\begin{equation}
\ket{\Omega} \equiv \ket{0}_{\text{P/A}}\otimes \ket{0}_{\text{spin}}\otimes \ket{0 ... 0}_{\text{momentum}}, 
\label{eq:creation-1}
\end{equation}
each set of $m>0$ qubits is able to store $2^m$ different values of the corresponding degree of freedom as binary numbers. The presence/absense qubit indicates whether the register is empty or occupied and it is used as control, one qubit is enough for spin, and the amount of momentum qubits depends on discretization.
The $a$ and $a^{\dagger}$ are written as tensor product of $\textit{set}$, $\textit{scrap}$ and control operators $\ketbrac{i}{j}\ket{k} = \delta_{jk}\ket{i}$:
\begin{equation}
a_{p,\eta}^{\dagger} \equiv \ketbrac{1}{0}\otimes \mathfrak{s}^{\dagger}_{\eta}\otimes \mathfrak{s}^{\dagger}_{p},\,\,\, a_{p,\eta} \equiv \ketbrac{0}{1}\otimes \mathfrak{s}_{\eta}\otimes \mathfrak{s}_{p},
\label{eq:creation-2}
\end{equation}
which initiate the corresponding qubits to the binary number that represents the value of the degree of freedom along with the presence qubit:
\begin{equation}
a_{p,\lambda}^{\dagger}\ket{\Omega} = \ket{1}\otimes\ket{\eta}\otimes\ket{p} = \ket{1\eta p}.
\end{equation} 
We use the abbreviated notation $a_{r}^{\dagger} =  \ketbrac{1}{0}\otimes \mathfrak{s}^{\dagger}_{r}$ to collectively denote all quantum numbers.

Some problems arise with states of several particles, the vacuum for \textit{up to} $n$ particles is the tensor product of $n$ single-particle vacua
\begin{equation}
\ket{\Omega} \equiv \ket{\Omega}_n\otimes...\otimes\ket{\Omega}_1,
\end{equation}
and the memory is considered to be filled from right to left, the first unoccupied register being found using projectors:
\begin{align}
\projector{n}{j} & \ket{\Omega}_n\otimes....\ket{\Omega}_{k+1}\otimes\ket{1\eta'p'}_k\otimes...\otimes\ket{1\eta p}_1\nonumber\\
 = & \,\delta_{j,k}\ket{\Omega}_n\otimes....\ket{\Omega}_{k+1}\otimes\ket{1\eta'p'}_k\otimes...\otimes\ket{1\eta p}_1,
\end{align}
which can be written in terms of the $set$, $scrap$ and control operators described earlier. To implement the appropiate particle statistics we introduce step-antisymemtrizers $\mathcal{A}_{j\leftarrow(j-1)}$, which antisymmetrize register $j$ with all the others, already assumed antisymmetric. In this way we can write the creation and annihilation operators
\begin{equation}
a^{(n)\dagger }_{r} = \sum^{n}_{j = 1}  \mathcal{A}_{j\leftarrow(j-1)} \cdot \projector{n-j}{0}\otimes \left(\ketbrac{1}{0}\otimes \mathfrak{s}^{\dagger}_{r}\right)_{j} \otimes \,\projector{j-1}{j-1},
    \label{def:nReg-bosoncreation}
\end{equation}
and 
\begin{equation}
a^{(n)}_{r} = \sum^{n}_{j=1} \projector{n-j}{0}\otimes \left(\ketbrac{0}{1}\otimes \scraps{r}\right)_{j} \otimes \,\projector{j-1}{j-1} \cdot\mathcal{A}_{j\leftarrow(j-1)},
    \label{def:nReg-bosonannihilation}
\end{equation}
with a `$\cdot$' to separate operators over the entire memory and each register acts on the number of registers indicated by its superscript $(n)$, $(n-j)$, etc.
\begin{align}
&\left\{ a^{(n)}_{r},a^{(n)\dagger}_{s}\right\}\   = \ \ \delta_{r,s}\left(\ketbrac{0}{0}\otimes\innerid\right)_{n}\otimes \sum^{n-1}_{j=0}\projector{n-1}{j}\ \  \nonumber \\
& +\  \ \mathcal{A}_{n\leftarrow n-1}\cdot \left(\ketbrac{1}{1}\otimes \mathfrak{s}^{\dagger}_{r}\mathfrak{s}_{s}\right)_n\otimes\projector{n-1}{n-1}\cdot \mathcal{A}_{n\leftarrow n-1} ,
\end{align}
with a non-canonical boundary term that vanishes as long as there is at least one empty register, and is equivalent to defining the boundary action of the creation operator as
\begin{equation}
a^{(n)\dagger}_r\ket{1\eta'p'}_n\otimes...\otimes\ket{1\eta p}_1 = 0.
\end{equation}
In the examples below the number of particles is kept fixed, so the boundary condition has no effect. Thus we can use Eq.~(\ref{def:nReg-bosoncreation}) and Eq.~(\ref{def:nReg-bosonannihilation}) to obtain expressions in terms of $\textit{set}$ and $\textit{scrap}$ operators that can be exponentiated and applied to the quantum memory. Details are to be found in \cite{galvez-viruet_2409_2024}; we here just state the results for a quantum memory of two registers: the time evolution operator for a kinetic-energy term such as the exponential of the first term of Eq.~(\ref{eq:Hamiltonian}) below can be written as
\begin{align}
\mathcal{U}_{11}&\left(\Delta t\right) = \projector{2}{0} +\projector{1}{0}\otimes\left(\ketbrac{1}{1}\otimes\mathfrak{U}_{11}\left(\Delta t\right)\right)_1\nonumber \\ & + \left(\ketbrac{1}{1}\otimes  \mathfrak{U}_{11}\left(\Delta t\right)\right)_{2}\otimes \left(\ketbrac{1}{1}\otimes  \mathfrak{U}_{11}\left(\Delta t\right)\right)_{1},
\label{eq:free-evolution}
\end{align}
where 
\begin{equation}
\mathfrak{U}_{11}(\Delta t)\equiv \exp\left[-i\Delta t \sum_{k} e_k\, \innerid\otimes\sets{k}\scraps{k}\right],
\label{def:free-evolution-onregister}
\end{equation}
adds to each register a phase corresponding to its kinetic energy $e_k$ and the spin operators have been summed to the identity `$\innerid$'. Eq.~(\ref{eq:free-evolution}) is the free-energy evolution operator (over the entire memory) and it defines how the operators over single registers, Eq.~(\ref{def:free-evolution-onregister}), should be applied. The implementation is completed once these exponentials are written in terms of basic gates, a decomposition that can be done using built-in functions of programming packages such as \textit{qiskit} \cite{qiskit2024}.

The exponential of the momentum exchange term of Eq.~(\ref{eq:Hamiltonian}) simplifies to
\begin{align}
\mathcal{U}_{22}&\left(\Delta t\right) = \projector{2}{0} + \projector{2}{1} + \left\{\ketbrac{1}{1}\otimes\ketbrac{1}{1}\right\}\left\{\mathfrak{U}_{22,2.1}(\Delta t)\right\}, 
\label{eq:momentum-exchange}
\end{align}
where the control operators act on the presence qubits of the first two registers. The register operator is now
\begin{equation}
\mathfrak{U}_{22,2.1}(\Delta t)= e^{-i\Delta t\sum_{q}\sum_{p,\eta_1;k,\eta_2}\lambda_{q}\left[\sets{r} \scraps{s} \,\otimes\, \sets{t}\scraps{u} + \sets{t} \scraps{u} \,\otimes\, \sets{r}\scraps{s}\right]},
    \label{eq:four-body-I-fermions-exp}
\end{equation}
where the $\textit{set}$ and $\textit{scrap}$ operators separately act on spin and momentum qubits:
\begin{align}
\sets{r}\scraps{s} & = \sets{\eta_1}\scraps{\eta_1}\otimes\sets{p+q}\scraps{p},\nonumber \\
\sets{t}\scraps{u} & = \sets{\eta_2}\scraps{\eta_2}\otimes\sets{k-q}\scraps{k},
\end{align}
so the spin operators can be again summed to the identity.

The total evolution operator is implemented using the Trotter formula
\begin{align}
\mathcal{U}(\Delta t) & = \left\{\mathcal{U}_{11}\left(\frac{\Delta t}{n_\text{Trotter}}\right)\mathcal{U}_{22}\left(\frac{\Delta t}{n_\text{Trotter}}\right)\right\}^{n_\text{Trotter}}\nonumber \\
& + \mathcal{O}\left(\frac{\Delta t}{n_\text{Trotter}}\right)^2,
\label{eq:trotter}
\end{align}
where we call the time step $\frac{\Delta t}{n_\text{Trotter}}$ \textit{Trotter interval}. The equation allows for the simulation of time evolution with a systematically improvable error.

\section{Antisymmetrization}
\label{sec:anti}
In the state praparation process or during the evolution of the quantum memory it is sometimes necessary to apply antisymmetrizers over arbitrary states, an operation that is not unitary. To circumvent the issue we impose an ordering to the memory registers using comparisons between binary numbers; we say that $\ket{1s_0 p_0}>\ket{1s_1p_1}$ if the binary representing the state $\ket{1s_0 p_0}$ is greater than that representing the state $\ket{1s_1p_1}$. This ordering can be used to disentangle the auxiliary qubits that make the operation feasible.

As an example, consider two scalar particles with two momentum qubits in which we codify three different momenta:
\begin{align}
\ket{00}& \rightarrow \text{None},\,\,\ket{01}\rightarrow\ket{p_0},\nonumber \\
\ket{10}& \rightarrow\ket{p_1},\,\,\ket{11}\rightarrow\ket{p_2},
\end{align}
in which the state $\ket{00}$ has been excluded for technical reasons. Adding the presence qubit in total gives three qubits per particle.

We first initiate the memory in the following superposition (note that $p_1$, whose binary number is larger than that of $p_0$, is written in the second register):
\begin{equation}
\ket{\phi_0} = \frac{1}{\sqrt{2}}\left(\ket{\Omega}_2\ket{1p_0}_1+\ket{1p_1}_2\ket{1p_0}_1\right),
\end{equation}
two auxiliary qubits are now added, the first one associated to the largest state (as defined earlier) if there are at least two active registers:
\begin{align}
\ket{\phi_1} & = \frac{1}{\sqrt{2}}\left(\ket{00}_{ax}\ket{\Omega}_2\ket{1p_0}_1\right.\nonumber \\
& \left.+\ket{10}_{ax}\ket{1p_1}_2\ket{1p_0}_1\right),
\end{align}
the auxiliary register is antisymmetrized, and {\tt SWAP} operations and a phase flip are applied to the registers controlling on whether the second auxiliary qubit is $1$: 
\begin{align}
\ket{\phi_2} & = \frac{1}{\sqrt{2}}\ket{00}_{ax}\ket{\Omega}_2\ket{1p_0}_1\nonumber \\
& +\frac{1}{2}\left(\ket{10}_{ax}\ket{1p_1}_2\ket{1p_0}_1-\ket{01}_{ax}\ket{1p_0}_2\ket{1p_1}_1\right),
\label{eq:antisy-state-entangled}
\end{align}
the auxiliary qubit is now disentangled with an algorithm called \textit{Locate the Largest}, which points, on an auxiliary register, to the position of the largest momenta in a quantum memory with at least two registers (this algorithm is described in \cite{galvez-viruet_2409_2024}). Applying the algorithm we get
\begin{align}
\ket{\phi_3} & = \ket{00}_{ax}\left[\frac{1}{\sqrt{2}}\ket{\Omega}_2\ket{1p_0}_1\right.\nonumber \\
& \left.+\frac{1}{2}\left(\ket{1p_1}_2\ket{1p_0}_1-\ket{1p_0}_2\ket{1p_1}_1\right)\right],
\label{eq:antisy-state}
\end{align}
so the memory is antisymmetrized and the auxiliary qubits unentangled. This procedure has been validated on a quantum computer, both via classical simulation with qiskit \textit{aer} \cite{qiskit2024} and with the \textit{ibm brisbane} chip, accesible by the IBM cloud. 

\subsection{Numerical demonstration} 
\begin{figure}[t!]
\begin{center}
\includegraphics[scale=0.4]{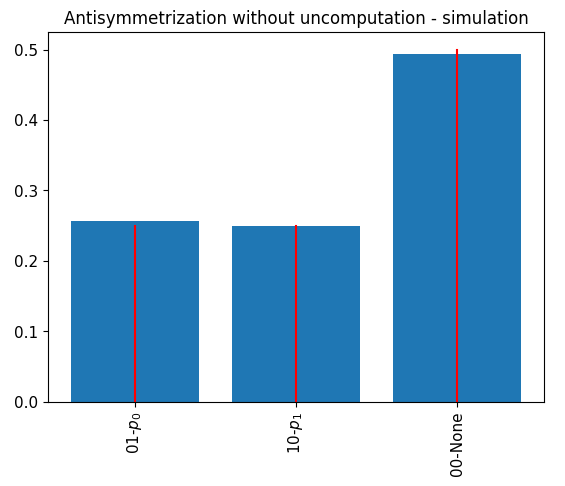}
\includegraphics[scale=0.4]{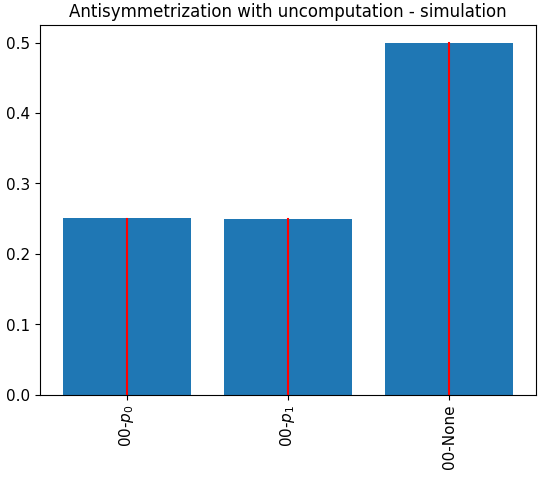}
\end{center}
\vspace{-0.5cm}
\caption{Measurement of the auxiliary and second registers of the antisymmetrization example  Sec.~\ref{sec:anti}, binary digits before the dash correspond to auxiliary qubits. The circuits have been executed with the \textit{qiskit aer} simulator.\label{fig: antisymmetrization-sim}}
\end{figure}

\begin{figure}[t!]
\begin{center}
\includegraphics[scale=0.4]{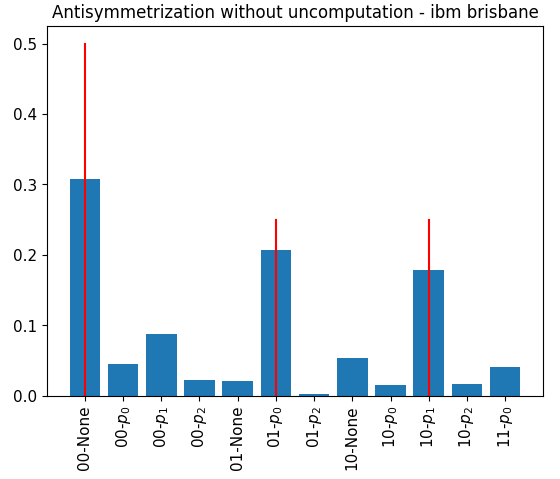}
\includegraphics[scale=0.4]{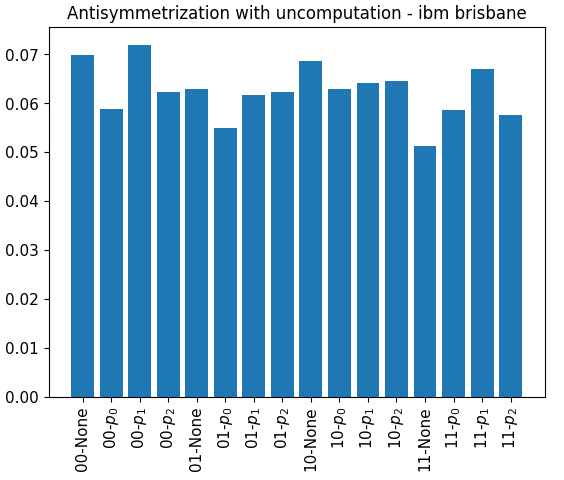}
\end{center}
\vspace{-0.5cm}
\caption{Measurement of the auxiliary and second registers of the antisymmetrization example Sec.~\ref{sec:anti}, binary digits before the dash correspond to auxiliary qubits. The histograms represent quasi-probabilities from the IBM quantum chip \textit{ibm brisbane}, see the text for futher details.\label{fig: antisymmetrization-real}}
\end{figure}

Figs.~(\ref{fig: antisymmetrization-sim}-\ref{fig: antisymmetrization-real}) are the results of measuring the momentum qubits of the second register of Eq.~(\ref{eq:antisy-state}), along with the auxiliary register, following the antisymmetrization process. In the figures, binary digits preceding the dash symbol correspond to auxiliary qubits.
The histograms of Fig.~(\ref{fig: antisymmetrization-sim}) are the outcomes of a quantum circuit simulation performed using IBM's Qiskit \textit{aer} simulator, both with and without the final unentangling step. A similar experiment was conducted on the IBM quantum chip \textit{ibm brisbane}, with 5000 shots and an error mitigation procedure applied to obtain quasi-probabilities, see Fig.~(\ref{fig: antisymmetrization-real}). The circuits without the uncomputation step (that is, with result Eq.~(\ref{eq:antisy-state-entangled})), have a depth of approximately 300 layers (first histogram), and it shows a strong resemblance to the simulator results, Fig~(\ref{fig: antisymmetrization-sim}). However, incorporating the unentangling steps to obtain Eq.~(\ref{eq:antisy-state}) increases the circuit depth to around 2000 layers, which exceeds the current capabilities of available quantum computers. The red vertical lines are the exact probabilities, momenta are measured to be in state $\textit{None}$ (corresponding to an empty register) with probability $1/2$, while momenta $p_0$ or $p_1$ are measured with probabilities $1/4$ each.

\section{Momentum exchange simulation}
\label{sec:Hamiltonian}
We now simulate the behaviour of two electrons under a positively charged background in two dimensions. We first state the Hamiltonian, closely following the discussion of Chap. 1, Sec. 3 of \cite{fetter_quantum_2003}. The operator is divided as follows
\begin{equation}
\hat{H} = \hat{H}_{el} + \hat{H}_{b} + \hat{H}_{el-b},
\end{equation}
where $\hat{H}_{el}$ and $\hat{H}_{b} $ are the Hamiltonians for the electrons and the background and $\hat{H}_{el-b}$ that of the interaction among the two. Assuming an innert background and a spatially uniform system, $\hat{H}_{b}$ and $\hat{H}_{el-b}$ can be demonstrated to be pure c-numbers. Thus we focus on the electron part, which in first quantization reads
\begin{equation}
\hat{H}_{el} = -\sum^{N_e}_{i} \frac{\hbar^2 \nabla^2_i}{2m}+\frac{1}{2}e^2\sum^{N_e}_{i\neq j}\frac{e^{-\mu |r_i-r_j|}}{|r_i-r_j|},
\end{equation}
where the first term is the kinetic energy $T$ and the second the potential energy $V$; $N_e$ is the number of electrons, $p_i$ their momentum, $m$ their mass, $e$ their charge and $r_i$ their positions. $\mu$ is a regulator needed to make the potential matrix elements finite, the pure c-numbers mentioned earlier also depend on $\mu$.

The passage to second-quantization is realized through the following equation
\begin{align}
\hat{H}_{el} & = \sum_{rs} a^{\dagger}_{r}\bra{r}T\ket{s}a_s \nonumber \\
& + \frac{1}{2}\sum_{rstu} a^{\dagger}_r a^{\dagger}_s \bra{rs}V\ket{tu}a_u a_t,
\label{eq:second-quantized}
\end{align}
where $r$, $s$, ... are to be understood as generic indices encompasing both the momentum and spin degrees of freedom, i.e. $a_r \equiv a_{k_1,\eta_1}$, $\ket{r}\equiv \ket{k_1,\eta_1}$, etc. The operators fulfill the anticommutation relations 
\begin{equation}
\left\{a^{\dagger}_{k_1,\eta_1},a_{k_2,\eta_2}\right\} = \delta_{k_1,k_2}\delta_{\eta_1,\eta_2}, 
\end{equation}
as we consider a discrete momentum values. To evaluate the matrix elements we approximate the discrete sums over position  by integrals, normalized with a total surface $S$, $\sum_x\rightarrow\frac{1}{S}\int d^2x$ and take as basis functions plane waves with periodic boundary conditions $k_i = \frac{2\pi n_i}{\sqrt{S}}$. A characteristic length $r_0 = \sqrt{\frac{S}{N\pi}}$ is also introduced to define dimensionless quantities, denoted with bars. The final Hamiltonian reads, 
\begin{align}
\hat{H} & = \sum_{\bar{k},\eta} e_k \, a^{\dagger}_{\bar{k}\eta}a_{\bar{k},\eta} \nonumber\\
&  +\sum_{\substack{\bar{k}_1,\bar{k}_2\\ \eta_1,\eta_2}}\sum_{\bar{q}\neq (0,0)}\lambda_q \,a^{\dagger}_{\bar{k}_1+\bar{q},\eta_1}a^{\dagger}_{\bar{k}_2-\bar{q},\eta_2}a_{\bar{k}_2,\eta_2}a_{\bar{k}_1,\eta_1},
 \label{eq:Hamiltonian}
\end{align}
where the condition $q\neq(0,0)$ is a consecuence of regularization and
\begin{equation}
e_k = \frac{|\bar{k}|^2\,E_0 }{2 r_s^2},\,\,\,\lambda_q = \frac{E_0}{r_s\,N_e\,|\bar{q}|},
\label{eq:Hamiltonian-constants}
\end{equation}
with $E_0 = e^2/a_0 = 340$ eV, $a_0$ the Bohr radius and $r_s = r_0/a_0$ a free parameter.

\subsection{Implementation}
\label{subsec:implementation}
\begin{figure}
\begin{center}
\includegraphics[scale=0.13]{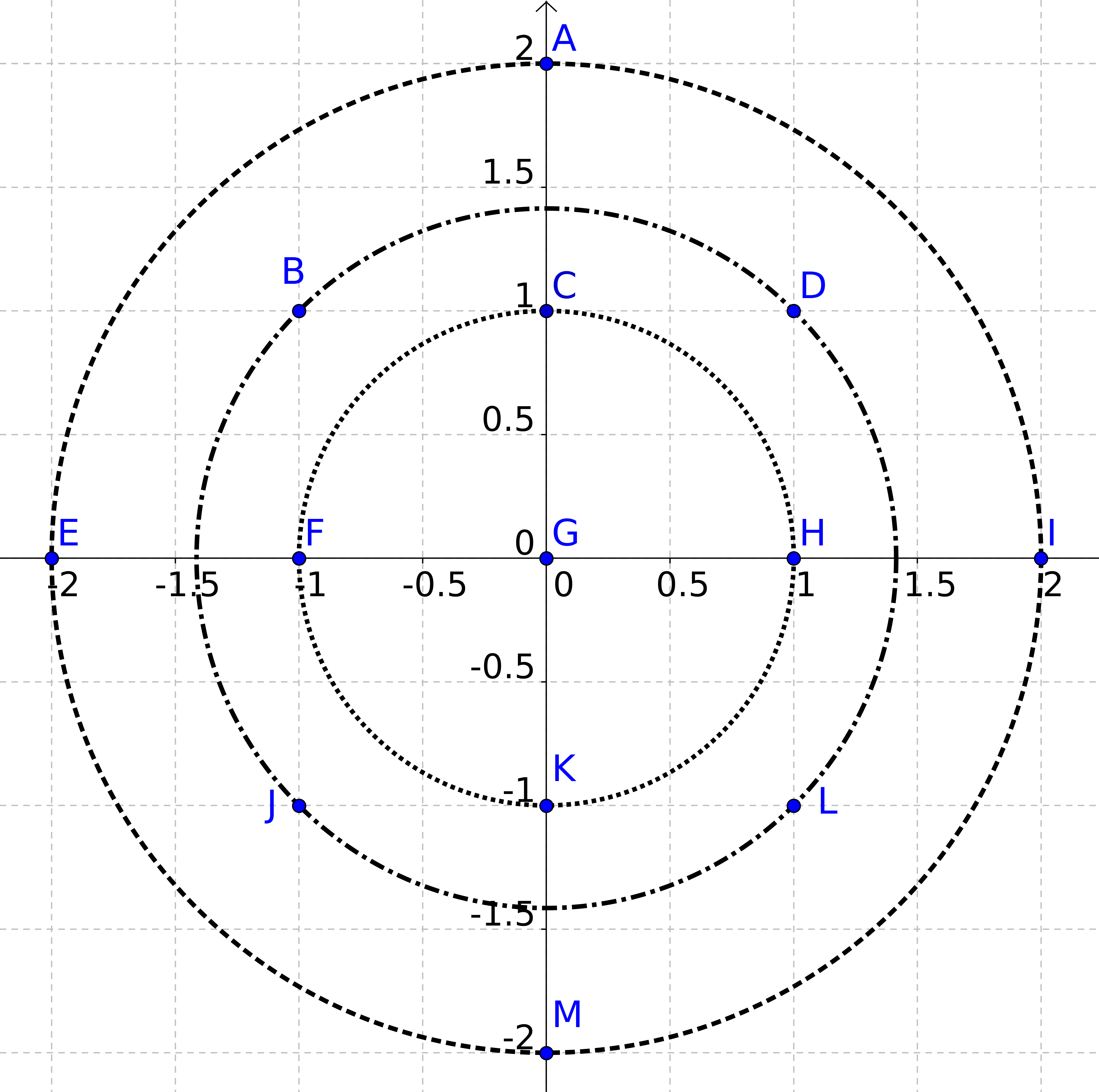}
\end{center}
\caption{Representation of the momentum lattice: Points lying on the same circunference have equal kinetic energy. Discretization $a)$ corresponds to points $\left\{C,F,G,H,K\right\}$, while discretization $b)$ includes all.  \label{fig:lattice}}
\end{figure}

The codification is now applied to the Hamiltonian Eq.~(\ref{eq:Hamiltonian}). We consider two particles and two different discretizations with different qubit counts:
\begin{table}[b!]
\centering
\begin{tabular}{|c|c|c|}
\hline
\textbf{Qubit states}          & \textbf{Lattice label} & \textbf{Physical state}              \\ \hline
$\left|000\right\rangle$ & None          & None                        \\ \hline
$\left|001\right\rangle$ & C             & $\left|(0,1)\right\rangle$  \\ \hline
$\left|010\right\rangle$ & F             & $\left|(-1,0)\right\rangle$ \\ \hline
$\left|011\right\rangle$ & G             & $\left|(0,0)\right\rangle$  \\ \hline
$\left|100\right\rangle$ & H             & $\left|(0,1)\right\rangle$  \\ \hline
$\left|101\right\rangle$ & K             & $\left|(0,-1)\right\rangle$ \\ \hline
\end{tabular}
\caption{Table of codification $a)$ with correspondence between qubit states and momentum states $\bar{k} = \sqrt{\frac{4\pi}{N_e}}(i,j)$. The qubit state $\ket{000}$ does not represent any physical state.\label{tab:codification}}
\end{table}
\begin{itemize}
\item[$a)$] 1 presence qubit + 1 spin qubit + 3 qubits to represent 5 values of momenta: 5 qubits per particle
\item[$b)$] 1 presence qubit + 1 spin qubit + 4 qutbis to represent 13 values of momenta: 6 qubits per particle
\end{itemize}
Momenta are arranged in the 2D symmetric lattice of Fig.~\ref{fig:lattice}, where discretization $a)$ corresponds to points $\left\{C,F,G,H,K\right\}$ and discretization $b)$ includes all.

Once physical states have been mapped to qubit states, the \textit{set} and \textit{scrap} operators which enter Eq.~(\ref{def:free-evolution-onregister}) and Eq.~(\ref{eq:four-body-I-fermions-exp}) can be written in terms of Pauli matrices $I$, $X$, $Y$, $Z$, for example
\begin{equation}
\sets{(0,1)} = \ket{001}\bra{000} = \frac{I+Z}{2}\otimes\frac{I+Z}{2}\otimes \frac{X-iY}{2},
\end{equation}
the exponentials of the sum of these tensor products can then be decomposed in terms of basic gates and applied to a quantum circuit using the built-in qiskit's function \textit{HamiltonianGate}; finally, the Trotter formula Eq.~(\ref{eq:trotter}) is used to control the errors comming from separate exponentiation of the free and exchange terms of the Hamiltonian Eq.~(\ref{eq:Hamiltonian}). The operator thus obtained is used to evolve, as an example, the two electron state
\begin{align}
\ket{\Psi}_\text{init} = &  \frac{1}{\sqrt{2}}\left(\ket{10}_\text{aux}\ket{\uparrow p_0}_2\ket{\downarrow p_0}_1\right.\nonumber \\
& \left.- \ket{01}_\text{aux}\ket{\downarrow p_0}_2\ket{\uparrow p_0}_1\right),
\label{eq:initial-state}
\end{align}
which is obviously not an eigenstate of the Hamiltonian Eq.~(\ref{eq:Hamiltonian}); the probabilities thus change with time, and we keep track of that of measuring $p_0$ on the first register. The auxiliary qubits are not uncomputed because the number of particles does not change and no further antisymmetization is needed. 
\subsection{Results}
To carry out the simulations we used the qiskit's \textit{aer} simulator with $N_e = 2$ for two electrons and $r_s = 30$ in Eq.~(\ref{eq:Hamiltonian-constants}) for the kinetic and momentum exchange terms to be comparable in size.
\begin{figure}[t!]
\begin{center}
\includegraphics[scale=0.4]{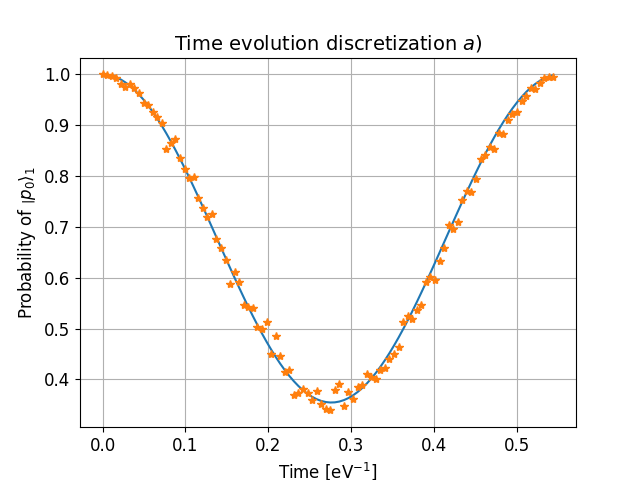}
\includegraphics[scale=0.4]{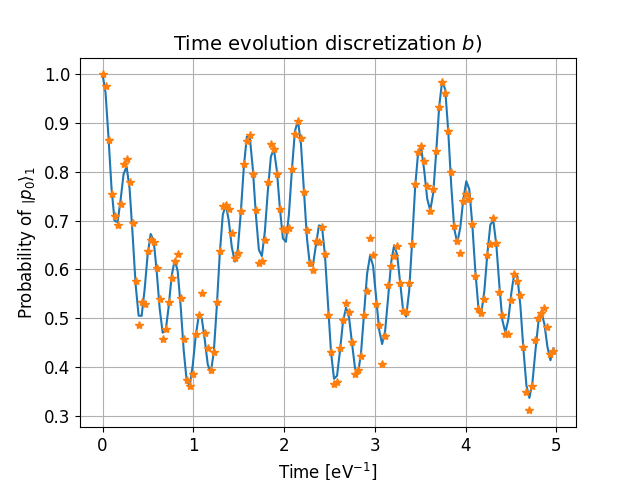}
\caption{Probability of measuring momentum $p_0$ in the first register as a function of time for discretizations $a)$ (up) and $b)$ (down). Orange star-like points are obtained by the \textit{qiskit aer} simulator of quantum circuits, while the blue line is the result of direct numerical exponentiation of the Hamiltonian matrix by conventional means.  \label{fig:time-evolution}}
\end{center}
\end{figure}

The results are shown in Fig.~(\ref{fig:time-evolution}), with 100 time points and a Trotter interval of 0.055 for $a)$ and 151 time points and a Trotter interval of 0.03 for $b)$; the number of circuit shots per time point is always 1000. The first plot confirms our expectations for $a)$, where we see a simple oscillatory behaviour, but the composite evolution of $b)$ we see in the second plot is not as easy to interpret. 
\begin{figure}[t!]
\begin{center}
\includegraphics[scale=0.4]{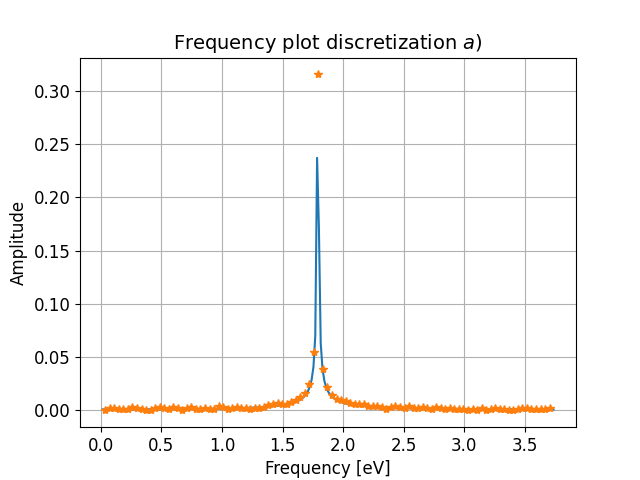}
\includegraphics[scale=0.4]{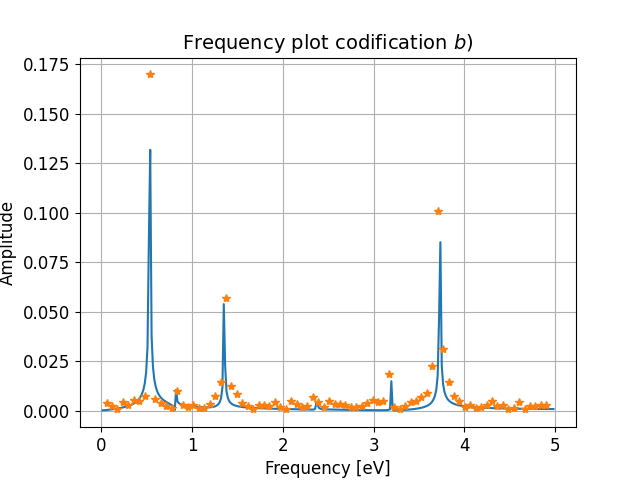}
\caption{Fourier transforms of the time-depending probaibility of measuring the state $p_0$ in the first register for discretizations $a)$ (up) and $b)$ (down). Orange star-like points are simulated, while the blue line is the result of direct numerical exponentiation.  \label{fig:frequency}}
\end{center}
\end{figure}

Fig.~(\ref{fig:frequency}) shows the frequency domain results, with 400 frequency points and a Trotter interval of 0.067 for $a)$ and 250 frequency points\footnote{Although it is evident that more data is needed to clearly distinguish the six peaks, the classical simulation of the 250 frequency points already took approximately 20 hours on a personal PC, with the time increasing as the number of frequencies squared.} and a Trotter interval of again 0.067 for $b)$, note that only half of the points really contributes to what we see in the figures, as the amplitudes are symmetric with respect to the ordinate axis. Numerical results for frequencies and periods are shown in Tab.~(\ref{table:results}). There we see a single frequency for discretization $a)$, as expected for the simple oscillatory behaviour of the time evolution plot. In contrast, there are $5$ clear peaks for discretization $b)$, but the exact solution shows another at approximately $2.4$ eV which is not resolved with only 250 frequency points; thus there are $6$ peaks, corresponding to an oscillation around $3$ states: the initial state, associated to point $G$ on Fig.~(\ref{fig:lattice}) and two other states, superpositions of points $\left\{C,F,H,K\right\}$ and $\left\{A,E,I,M\right\}$, the system does not oscillate to a superposition of points $\left\{B,D,J,L\right\}$.

\begin{table}
\centering
\begin{tabular}{l|c|c|}
\cline{2-3}
                            & \textbf{Frequency (eV)} & \textbf{Period (eV$^{-1}$)} \\ \hline
\multicolumn{1}{|l|}{$a_1$} & $1.79 \pm 0.04$         & $0.56 \pm 0.01$             \\ \hline
\multicolumn{1}{|l|}{$b_1$} & $0.54\pm 0.06$          & $1.9\pm 0.2$                \\ \hline
\multicolumn{1}{|l|}{$b_2$} & $0.84\pm 0.06$          & $1.2\pm 0.1$                \\ \hline
\multicolumn{1}{|l|}{$b_3$} & $1.37\pm 0.06$          & $0.73\pm 0.03$              \\ \hline
\multicolumn{1}{|l|}{$b_4$} & $3.17\pm 0.06$          & $0.32\pm 0.01$              \\ \hline
\multicolumn{1}{|l|}{$b_5$} & $3.71\pm 0.06$          & $0.269\pm 0.004$            \\ \hline
\end{tabular}
\caption{Frequencies and periods characterizing the evolution of the state Eq.~(\ref{eq:initial-state}) for discretizations $a)$ and $b)$ discussed in subsection \ref{subsec:implementation}. Errors correspond to the finite spacing between frequency values.\label{table:results}}
\end{table}

\section{Conclusions}
We have presented here a register-particle codification for a quantum computer, by which the unitary evolution operators are decomposed in terms acting on a small number of registers each time, and that can be implemented using standard quantum computation protocols. This has been applied to the antisymmetrization and time evolution of a two-register quantum memory and we have seen that unentanagling circuits are at present out of reach of this technology. Classical simulators can still be used to test and debug algorithms, but their applicability is limited to a low number of qubits, and therefore, to a very low number of particles.
Finally, we would like to note that with a few more qubits per register this example can be easily extended to incorporate other quantum numbers such as colour, flavour, etc.; with Eq.~(\ref{eq:second-quantized}) then promoting to a standard quark model, a step closer to QCD.

\section{Acknowledgements}
We would like to thank Felipe J. Llanes Estrada and María Gómez Rocha for a critical reading of the manuscript. Supported by grant PID2022-137003NB-I00 of the Spanish MCIN/AEI /10.13039/501100011033/; EU’s 824093
(STRONG2020); grant FPU21/04180 of the Spanish Ministry of Universities; Universidad Complutense de Madrid under research group 910309 and the IPARCOS
institute. 

\bibliographystyle{elsarticle-num}
\bibliography{MontBib_nonames}
\end{document}